\documentclass[12pt]{iopart}
\usepackage{iopams}
\usepackage[pdftex]{graphicx}
\expandafter\let\csname equation*\endcsname\relax
 \expandafter\let\csname endequation*\endcsname\relax
\usepackage{color,amsmath,amssymb,amsthm,times,graphics,graphicx,bbm,bm,enumerate}
\providecommand{\openone}{\leavevmode\hbox{\small1\kern-3.8pt\normalsize1}}

\begin{document}

\title{Universally Optimal Noisy Quantum Walks on Complex Networks}

\author{Filippo Caruso}
\address{LENS and Dipartimento di Fisica e Astronomia, Universit\`a di Firenze, I-50019 Sesto Fiorentino, Italy}
\address{QSTAR, Largo Enrico Fermi 2, I-50125 Firenze, Italy}
\ead{filippo.caruso@lens.unifi.it}

\begin{abstract}
Transport properties play a crucial role in several fields of science, as biology, chemistry, sociology, information science, and physics. The behavior of many dynamical processes running over complex networks is known to be closely related to the geometry of the underlying topology, but this connection becomes even harder to understand when quantum effects come into play. Here, we exploit the Kossakoski-Lindblad formalism of quantum stochastic walks to investigate the capability to quickly and robustly transmit energy (or information) between two distant points in very large complex structures, remarkably assisted by external noise and quantum features as coherence. An optimal mixing of classical and quantum transport is, very surprisingly, quite universal for a large class of complex networks. 
This widespread behaviour turns out to be also extremely robust with respect to geometry changes. These results might pave the way for designing optimal bio-inspired geometries of efficient transport nanostructures that can be used for solar energy and also quantum information and communication technologies.
\end{abstract}

\pacs{03.65.Yz,  05.60.Gg,  03.67.-a}

\maketitle

\section{Introduction}

The static properties and dynamical performances of complex network structures have been extensively studied in classical statistical physics \cite{Latora06}. In the last years there is a growing interest in fully understanding how these studies can be generalized when dealing with quantum mechanical systems, e.g. with microscopic and fragile quantum coherence affecting macroscopic and robust transport behavior.

Very recently, it has been found that quantum effects play a crucial role in the remarkably efficient energy transport phenomena in light-harvesting complexes \cite{engel07,lee,collini10,pani10,hildner,qbiobook,HP13}, also with a fundamental contribution of the external noisy environment \cite{MRLA2008,PH2008,CLFJ2008,CCDHP2009,CDCHP2010}. Indeed, an electronic exciton seems to behave as a quantum walker over these protein structures, exploiting quantum superposition and interference to enhance its capability to travel from an antenna complex, where light is absorbed, to the reaction center, where this energy is converted into a more available chemical form.
More in general, quantum walks \cite{aharonov} are becoming more and more popular recently, since they have potential applications, not only in energy transport \cite{blumen}, but also, for instance, in quantum information theory \cite{santha,farhi,childs,childs09,Kempe06,kendon,Andraca12} since they may lead to quantum algorithms with polynomial as well exponential speedups \cite{ambainis}, e.g. Grover search algorithm \cite{grover}, universal models for quantum computation \cite{childs13}, state transfer in spin and harmonic networks \cite{bose,ekert,PHE04,CHP2010}, noise-assisted quantum communication \cite{CHP2010}, and including also very recent proposals for google page ranking \cite{zueco12,paparo,GZL12,G12}. The generalization of classical random walks into the quantum domains had led to different variants, as, mainly, discrete-time quantum walks \cite{aharonov}, based on an additional action of a quantum `coin', and continuous-time quantum walks obtained, basically, by mapping the classical transfer matrix into a system Hamiltonian \cite{farhi}; notice, however, that these two types of walks are shown to be connected by a precise limiting procedure \cite{strauch}.

On top of that, this growing interest has been especially stimulated by the increasing number of fascinating and challenging experiments demonstrating the basis features of quantum walks. They have been implemented, for instance, with NMR \cite{du,ryan}, trapped ions \cite{schmitz,zahringer}, neutral atoms \cite{karski}, and several photonic schemes as waveguide structures \cite{perets}, bulk optics \cite{broome,white}, fiber loop configurations \cite{schreiber10,schreiber12,jeong}, and miniaturized integrated waveguide circuits \cite{peruzzo,owens,sansoni}. Most of these experiments included a single walker moving on a line and, only very recently, they have implemented optical quantum walks on a square lattice by laser pulses \cite{schreiber12} and single photons \cite{jeong}. Two-walker experiments on a line have been reported in Refs. \cite{peruzzo,owens,sansoni}, motivated also by the fact that the introduction of multiple walkers allows one to map a quantum walk on a line into higher dimensional lattices \cite{rohde}.

Therefore, given these first experimental attempts for higher-dimensional quantum walks, it turns out to be timely and interesting to investigate more in detail the transport features of quantum walkers over large complex networks. Furthermore, motivated by our previous results on noise-enhanced transfer of energy and information over light-harvesting complexes and communication networks \cite{CCDHP2009,CHP2010}, the additional mixing of quantum features with classical transport behaviour is worthwhile of being deeper understood when varying the underlying (high-dimensional) network topology. A recent study on the comparison of quantum and classical energy transfer, by noisy cellular automata, can be found in Ref. \cite{avalle}.
Fractal geometries have been also exploited to study the role of the topological 
structure on quantum transport \cite{Agliari08} and for Grover-based searching problems \cite{Agliari10} -- for a review on continuous-time quantum walks on complex networks see also Ref. \cite{blumen}. Concerning the role of decoherence in the mixing time of both discrete- and continuous-time quantum walks, especially for chains, cycles and hypercubes, a review paper is in Ref. \cite{kendon}. Moreover, in the context of light-harvesting phenomena, the role of geometry has been also investigated in terms of structure optimization \cite{H2011}, in presence of disordered systems \cite{MSLOR12,AMLZ12}, and to propose design principles for biomimetic structures \cite{SW2013}.

The outline of the paper is the following. In Sec. \ref{defs} we remind the standard formalism 
of complex networks, that are completely defined by their so-called adjacency matrices and 
spectral properties \cite{PVM2011}. In particular, one can define statistical measures of 
distances between nodes in terms of shortest path lengths and the corresponding maximum 
(or graph diameter), and local properties as clustering coefficient measuring the connectivity 
of each vertex. In Sec. \ref{dynam}, we describe the Kossakoski-Lindblad formalism of quantum 
stochastic walks, in terms of well defined master equations, generalizing classical (random) and 
quantum walks by including all possible transition elements from a vertex as given by the 
connectivity of the graph (adjacency matrix) \cite{Alan2010}. This formalism includes also, a
 special case, a model showing both classical and quantum walk behaviours as extreme cases 
 and the classical-quantum transition in the other regimes. Our figure of merit for the transport 
 efficiency of each graph is also defined. After these introductory sections, we investigate this 
 model for a family of large complex networks, comparing the corresponding transport 
 performances -- see Sec. \ref{results}. In particular, we find that the same mixing of classical 
 and quantum behaviours leads to the optimal transfer efficiency for all studied graphs. 
 This widespread behaviour is shown to be very robust when one modifies or deletes some 
 links of the structure. Finally, some conclusions and outlook are discussed in Sec. \ref{conclu}.

\section{The Model}
Here we briefly introduce the main notions of complex networks, together with some well-defined geometric measures, and then the formalism of quantum stochastic walks that has been considered in this paper in order to implement the transport dynamics over these graphs.

\subsection{Network topology: definitions and measures}
\label{defs}

The structure of a complex network or graph $G$ can be described by a pair $G=(V,E)$,
with $V(G)$ being a non-empty and finite set whose elements are
called vertices (or nodes) and $E(G)$ being a non-empty set of
unordered pairs of vertices, called edges (or links). Let us denote with $N$ the
number of nodes of the graph $G$, i.e. the number of elements
in $V(G)$, and with $L$ the number of links or elements of $E(G)$. A graph is completely defined by its adjacency matrix
$A$ as follows
\begin{equation}
  [A]_{i,j} := \left\{
\begin{array}{lll}
 1 && \mbox{if $\{V_i, V_j\}$ $\in$ $E(G)$} \\ \\
0 && \mbox{if $\{V_i, V_j\}$ $\notin$ $E(G)$.}
\end{array}
\right.
\end{equation}
Moreover, if $\{V_i, V_j\}$ $\in$ $E(G)$, the vertices $V_i$ and
$V_j$ are said to be adjacent or neighbor. The number of adjacent
vertices or neighbors to the vertex $V_i$ is denoted as $d_i$ and is
called the degree (or connectivity) of $V_i$, i.e. $d_i=\sum_{j=1}^{N} A_{ij}$. Notice that a graph
is said to be regular if each of its nodes has the same degree, i.e. $d_i=d$ for any $i$. In the following, we will consider only connected graphs, where there does exist always a path connecting any two nodes, since the unconnected subgraphs or isolated vertices do not play any role for the transport dynamics occurring over the rest of the network. Moreover, we restrict
to consider graphs without loops, i.e. without edges of the form
$\{V_i, V_i\}$, i.e. $A_{ii}=0$ for any $i$.

Although there are several measures of network topology in literature \cite{Latora06}, here we focus on the ones that are more related to our study. First of all, the number of hops from the node $i$ to $j$ with length $k$ equals to $A^k_{ij}$. Then, the shortest path length ${\cal L}_{ij}$ between $i$ and $j$ is the minimum number of steps (geodesic lengths) to go from node $i$ to node $j$, i.e.
\begin{equation}
{\cal L}_{ij}=\min k : \ A^k_{ij} > 0 \ \text{while} \ A^m_{ij} = 0 \ \text{for} \ m<k \; .
\end{equation}
The so-called characteristic path length $\cal{L}$ is hence defined as the average of the shortest paths between all possible pairs of nodes, i.e. ${ \cal L}=(1/\bar{L}) \sum_{i j} {\cal L}_{ij}$ with $\bar{L}=N(N-1)/2$ being the total number of node pairs. Furthermore, the largest ${\cal L}_{ij}$ is defined as the diameter $D$ of the graph, that is the largest distance (or longest, shortest path) between any two vertices of a graph. It corresponds also to the the lowest integer $k$ for which $(A^k)_{ij} \neq 0$ and $(A^m)_{ij} = 0$ if $m < k$ for each couple of nodes $i$ and $j$, i.e.
\begin{equation}
D=\max_{i,j} {\cal L}_{ij} \; .
\end{equation}
Note that for any graph $1 \leq D \leq N-1$.
Furthermore this quantity is closely related to the spectral properties of $A$. A well-known theorem in spectral theory of complex networks states, in fact, that the number of distinct eigenvalues $\pi$ of the adjacency matrix $A$ is at least equal to $D+1$, i.e. $\pi \geq D+1$ \cite{PVM2011}. On one side, the network with the smallest diameter is the complete or fully connected (FC) graph, where one has $D=1$, since all pairs of nodes are connected through a link. Indeed, the latter is the only graph whose adjacency matrix $A$ has only two different eigenvalues ($N-1$ with degeneracy $1$ and $-1$ with degeneracy $N-1$; note that the trace of $A$, and then the sum of its eigenvalues, has to be always $0$) and the bound above between $D$ and $\pi$ is tight. On the other side, for a given number of vertices $N$, the linear chain is the topology with the largest diameter ($D=N-1$).

Finally, another measure of the graph connectivity, known as clustering coefficient $\cal C$, quantifies how well the neighborhood of a node is connected \cite{WS98}. Given a node $i$, let us define $G_i$ as its neighborhood, i.e. the graph represented by the set of neighbors of the vertex $i$ and the relative interconnecting links. Then ${\cal C}_i$ is the local clustering coefficient of the node $i$ and is defined as ${\cal C}_i = 2 e_i /(d_i (d_i-1))$, with $e_i$ being the number of links in $G_i$ and $d_i$ the degree of the node $i$, i.e. it is the ratio of the number of links in $G_i$ ($e_i$) over its maximum possible number ($d_i (d_i-1)/2$). The (global) cluster coefficient $\cal C$ of the graph $G$ is then given by the average of ${\cal C}_i$ over all sites $i$, i.e.
\begin{equation}
{\cal C} = \frac{1}{N} \sum_{i=1}^{N} \frac{2 e_i}{d_i (d_i -1)} \; ,
\end{equation}
and it is always in the range $[0,1]$.
\subsection{Transport dynamics: quantum stochastic walks}
\label{dynam}

Once we have introduced the topological structure of our model, we need to specify the corresponding dynamics. To start with, let us remind that a random walk is usually defined as a time-discrete process where at each step the walker jumps between two connected nodes of the graph $G$ with some probability described by the transition matrix $T=\{T_{ij}\}$ \cite{weiss}. Usually, one has $T_{ij}=A_{ij}/{d_i}$ for classical random walks, that is indeed called as random walk normalized Laplacian matrix. More specifically, the Laplacian matrix $L$ is $L=A-D$, where $D$ is the diagonal matrix of the vertex degrees, i.e. $D=\{d_i\}$, while $T=D^{-1} A$. Given the occupation probability distribution $\vec{q}_t \equiv \{q_i^{(t)}\}$ of the walker over the nodes $V_i$ at a time $t$, the distribution at time $t+1$ is simply given by $\vec{q}_{t+1}= T \vec{q}_{t}$. The time-continuum version of such classical random walk (CRW) dynamics is then
\begin{equation}
\label{CRW}
\frac{d}{dt} {\vec{q}}= (T-\openone) \vec{q} \; .
\end{equation}

In both cases, the system ends up with a stationary (unique) distribution $\bar{q}_{i}=d_i/(2 N)$, that is the left-eigenvector of $T$ associated to the eigenvalue equal to $1$. The convergence rate towards the steady state $\bar{q}$, also known as mixing rate $\tau_{mix}$, is proportional to the so-called spectral gap of the graph $G$, that is the difference between the absolute values of the two largest eigenvalues of $A$ ($\lambda_1$, $\lambda_2$), i.e. $\tau_{mix} \propto (\lambda_1-\lambda_2)$ \cite{PVM2011}. In other words, the larger the spectral gap is, the faster the walker converges to $\bar{q}$. Note that one always has $\lambda_1-\lambda_2 \leq N$ and the bound is achieved for the FC graph. It is well known that well-connected graphs have small diameters and large spectral gaps, implying that the mixing rate $\tau_{mix}$ is larger (fast classical random walks) for graphs with larger $D$. There do exist several bounds showing that the mixing rate $\tau_{mix}$ does monotonically increase with $D$, apart from some constants.

In order to study the transport properties of such networks including also quantum coherence effects, we use the general framework of continuous-time quantum stochastic walks (QSW) \cite{Alan2010}, based on the Kossakowski-Lindblad master equation, allowing us to interpolate between classical random walks (CRW) and quantum walks (QW) \cite{Kempe06}. More specifically, the evolution of our initial state, represented by the density operator $\rho$, follows:
\begin{equation}
\label{QSRWs}
\frac{d \rho} {dt} = - (1-p) i [H, \rho] + p \sum_{i j} \left(L_{i j} \rho {L}^{\dagger}_{i j} - \frac{1}{2} \{{L}^{\dagger}_{i j} L_{i j} , \rho \} \right) \; ,
\end{equation}
where $H$ is the Hamiltonian describing the quantum coherent dynamics, while the operators $L_{i j}$ are responsible for the irreversibility. Note the parameter $p \in [ 0 , 1]$ quantifying the interplay between coherent (unitary) dynamics and incoherent (irreversible) one. In order to match with the limit of usual CRWs by choosing $p=1$, we assume $L_{i j} = T_{ij} | i \rangle \langle j |$ with $| i \rangle$ being the site basis, recovering Eq. (\ref{CRW}) for the diagonal elements of $\rho$, i.e. $\vec{q}_t \equiv \{\rho_{i i} (t)\}$. On the other side, $p=0$ corresponds to the pure QW master equation, where we choose $H=A$. This corresponds to a simplified model of a system (for instance, a light-harvesting complex), where all the energies and couplings are the same (homogeneous graphs) \cite{note}. For simplicity, we neglect the presence of losses and we restrict to the case where only one excitation is present in the network. Note that this formalism has been also used in a different context to propose a more efficient quantum navigation algorithm to rank elements of large (google page) networks \cite{zueco12,paparo}. 
To study the transfer efficiency of the QSW over the complex network $G$, on the right-side of Eq. (\ref{QSRWs}) we add another Lindblad term $L_{N+1} = \Gamma \ [ \ 2 |N+1 \rangle \langle N| \ \rho \ |N \rangle \langle N+1| - \{|N \rangle \langle N|,\ \rho\} ]$, with $\Gamma$ being the irreversible transfer rate from the site $N$ of the graph into some external node $N+1$ (trapping site or sink) where the energy is continuously and irreversibly stored. Hence, the transfer efficiency of the graph will be measured by \cite{PH2008,CCDHP2009,CDCHP2010,CHP2010}
\begin{equation}
{\cal E}(p,t) = 2 \Gamma \int_{0}^t\rho_{NN}(p,t')\mathrm{d}t' \; .
\end{equation}
Hence, our figure of merit will be:
\begin{equation}
{\cal E}(p) \equiv {\cal E}(p,\bar{t}) \; \text{with} \; \bar{t}: \; \exists \bar{p} \in [0,1] \; \text{with} \; {\cal E}(\bar{p},\bar{t}) \simeq 1 \; ,
\end{equation}
usually corresponding to take $\bar{t} \gg N ||A_{ij}||^{-1}$ with $||A_{ij}||$ being the strength of the coupling rates, i.e. considering the trapping site population after the initial transient behavior.
\section{Results}
\label{results}
In the following we will investigate the model described above for a large class of complex networks, including up to thousands of nodes, comparing the relative transfer efficiency ${\cal E}(p)$.
\begin{figure}[t]
\centerline{\includegraphics[width=.7\textwidth]{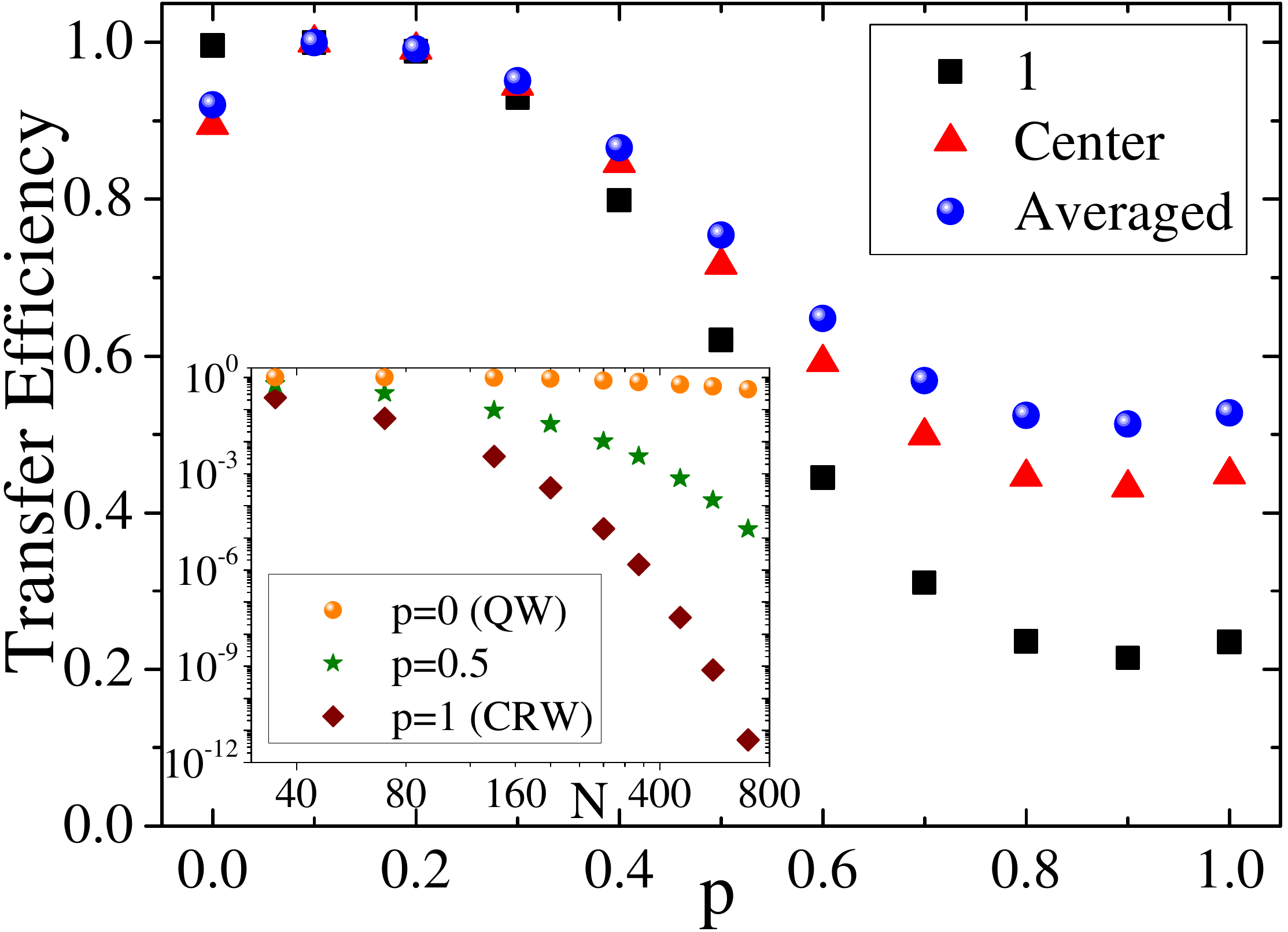}}
\caption{Transfer Efficiency ${\cal E}(p)$ as a function of $p$ for QSWs on a linear chain of $N=35$ nodes. Different initial conditions are considered, while the sink is always connected to the site $35$. Note that ${\cal E}(p)$ appears to be rising again towards $p=1$, perhaps because the coherent dynamics tends to be frozen (quantum Zeno effect) in presence of very strong irreversibility but the transport is recovered again at the classical limit. However, this is not always the case for other networks -- see, for instance, Fig. \ref {fig12} for random graphs.
Inset: ${\cal E}(p,\bar{t})$ vs. $N$, at time $\bar t$ being linearly proportional to $N$, in the case of a walker starting from site $1$. 
An asymptotic power law behaviour is observed but with an exponent depending on the value of $p$.}\label{fig1}
\end{figure}
\begin{figure}[b]
\centerline{\includegraphics[width=.43\textwidth]{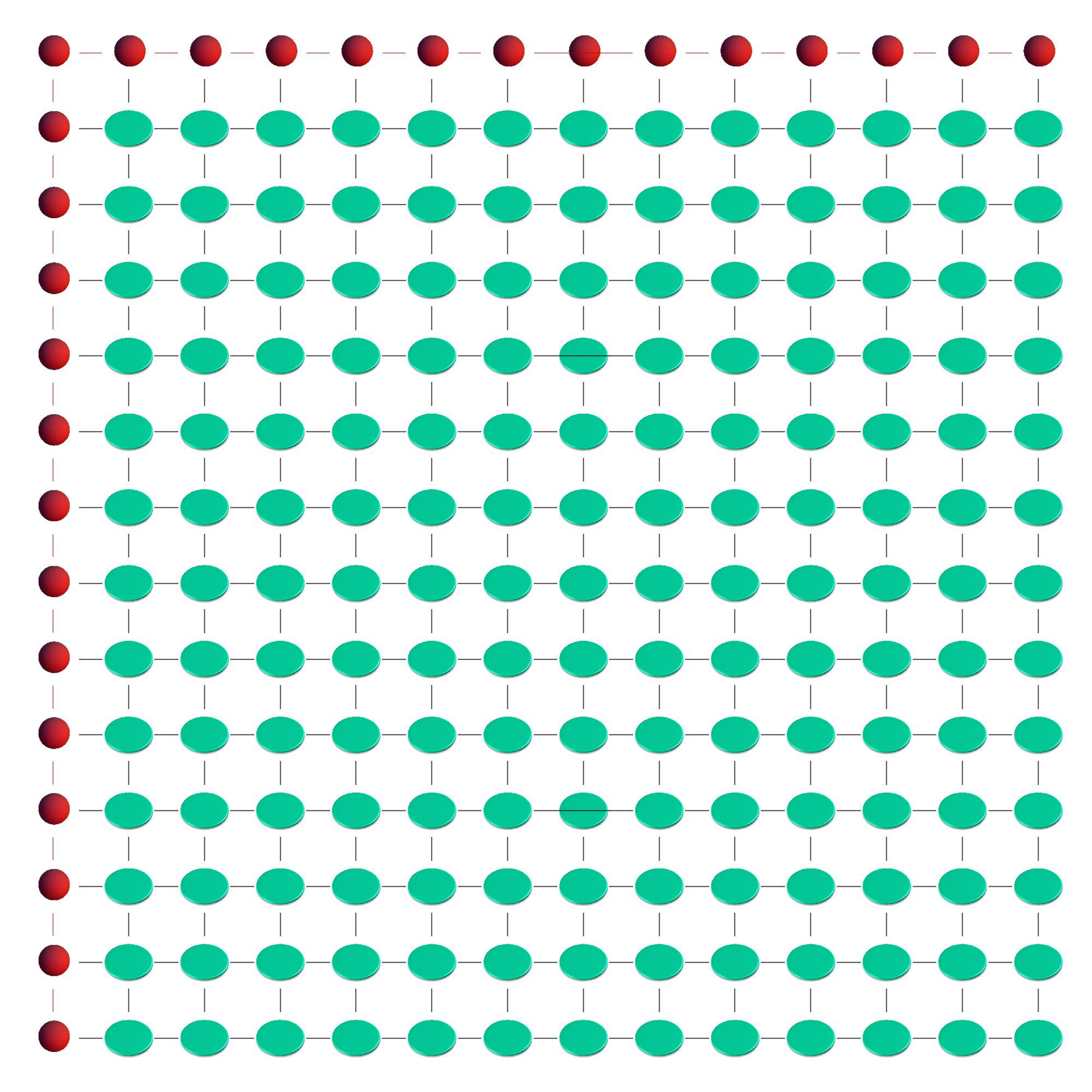}}
\caption{Square lattice of $N=196$ nodes. The red ones 
represent one of the shortest paths connecting two opposite vertices,
that are the injecting site ($1$) and the one ($196$) connected to the external sink, respectively.}\label{fig2}
\end{figure}
\subsection{Regular graphs: chains and square lattices}
The simplest geometry, to start with, is represented by a linear chain connecting node $1$ to node $N$ that is linked to the sink ($D=N-1$). As shown in Fig. \ref{fig1}, when the excitation initially is on site $1$ and in absence of disorder (homogeneous chain), the quantum limit ($p=0$) provides the optimal transport efficiency, in agreement with previous results in Refs. \cite{PH2008,CCDHP2009}.
However, this is not the case when the initial site is not located opposite to the sink \cite{KG12}.
Indeed, if it is on the center site of the chain, the efficiency becomes optimal at the value of $p \sim 0.1$. 
This does hold also when we average ${\cal E}(p)$ over any possible initial site -- see Fig. \ref{fig1} for a chain with $N=35$.
The same optimal value is also found for longer chains, although the scaling of ${\cal E}(p,\bar{t})$ with $N$ (at time $\bar t$ linearly proportional to $N$) does depend on $p$ and shows a stronger $N$--dependence (i.e. less robustness) in the classical limit
-- see inset of Fig. \ref{fig1}. 

\begin{figure}[t]
\centerline{\includegraphics[width=.65\textwidth]{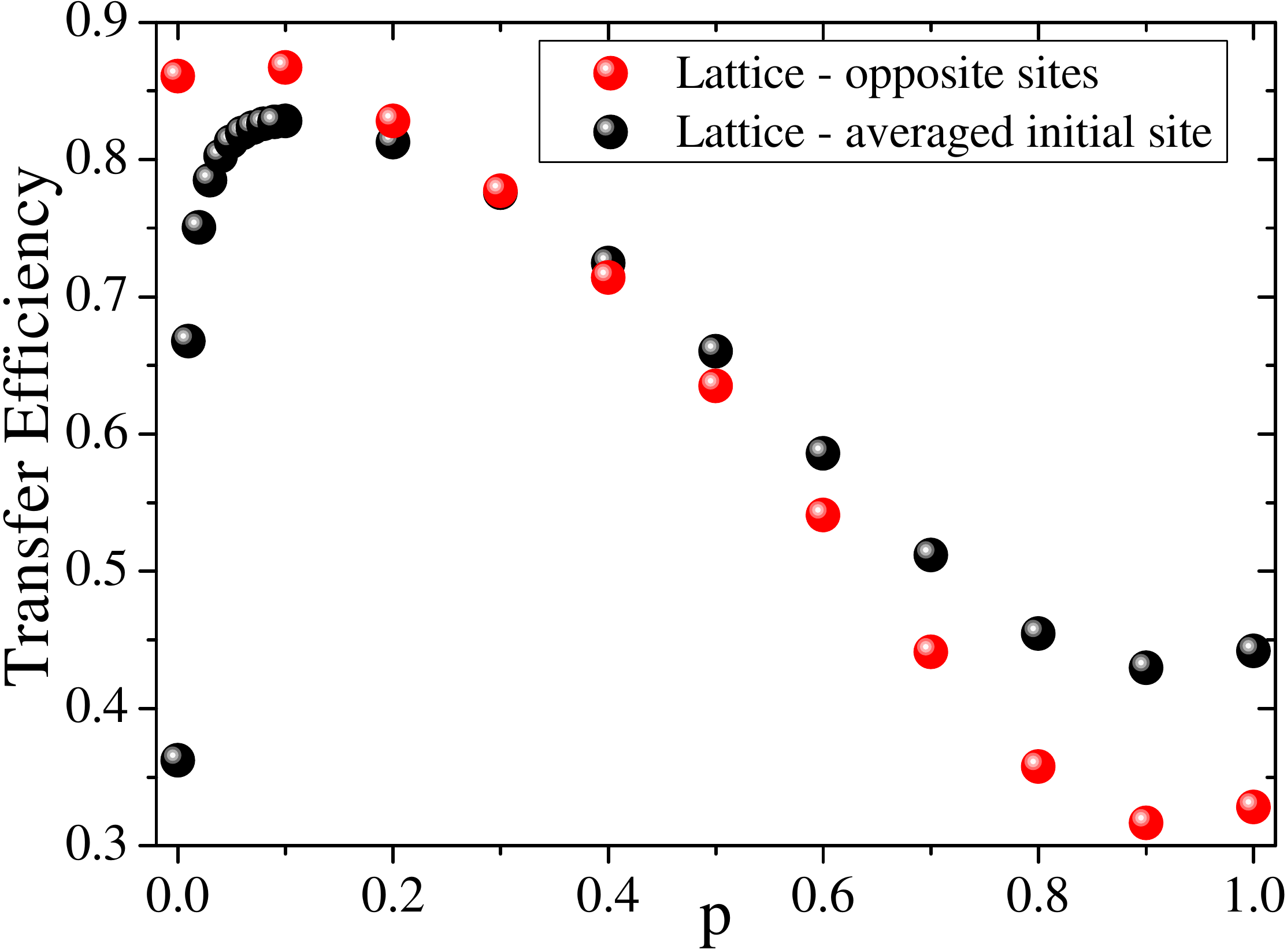}}
\caption{Transfer Efficiency ${\cal E}(p)$ vs. $p$ for QSWs on a square lattice of $N=196$ nodes, where the excitation is initially localized on a single site (1), while the sink is always at the site $196$. 
The averaging over all possible initial sites is also included.}\label{fig3}
\end{figure}

A $m \times m$ regular square lattice is a grid graph whose $N=m^2$ vertices are on a square grid ($D=2m-2$) and are linked only by
nearest neighbor edges -- see Fig. \ref{fig2}. It is considered as a regular graph since each vertex has degree $4$ (excluding the
boundary nodes). This network is typically characterized by high values of $\cal L$ but small clustering coefficients $\cal C$. 
As for linear chains, the optimal transport occurs for an intermediate mixing of quantum and classical effects, particularly for $p \sim 0.1$ -- see Fig. \ref{fig3}. 
Here and in the following, the decreasing behavior of ${\cal E}(p)$, for $p$ larger than the optimal one, is intuitively expected because of quantum Zeno effects suppressing the transport dynamics.

\subsection{Small-World topologies}
\begin{figure}[t]
\centerline{\includegraphics[width=.5\textwidth]{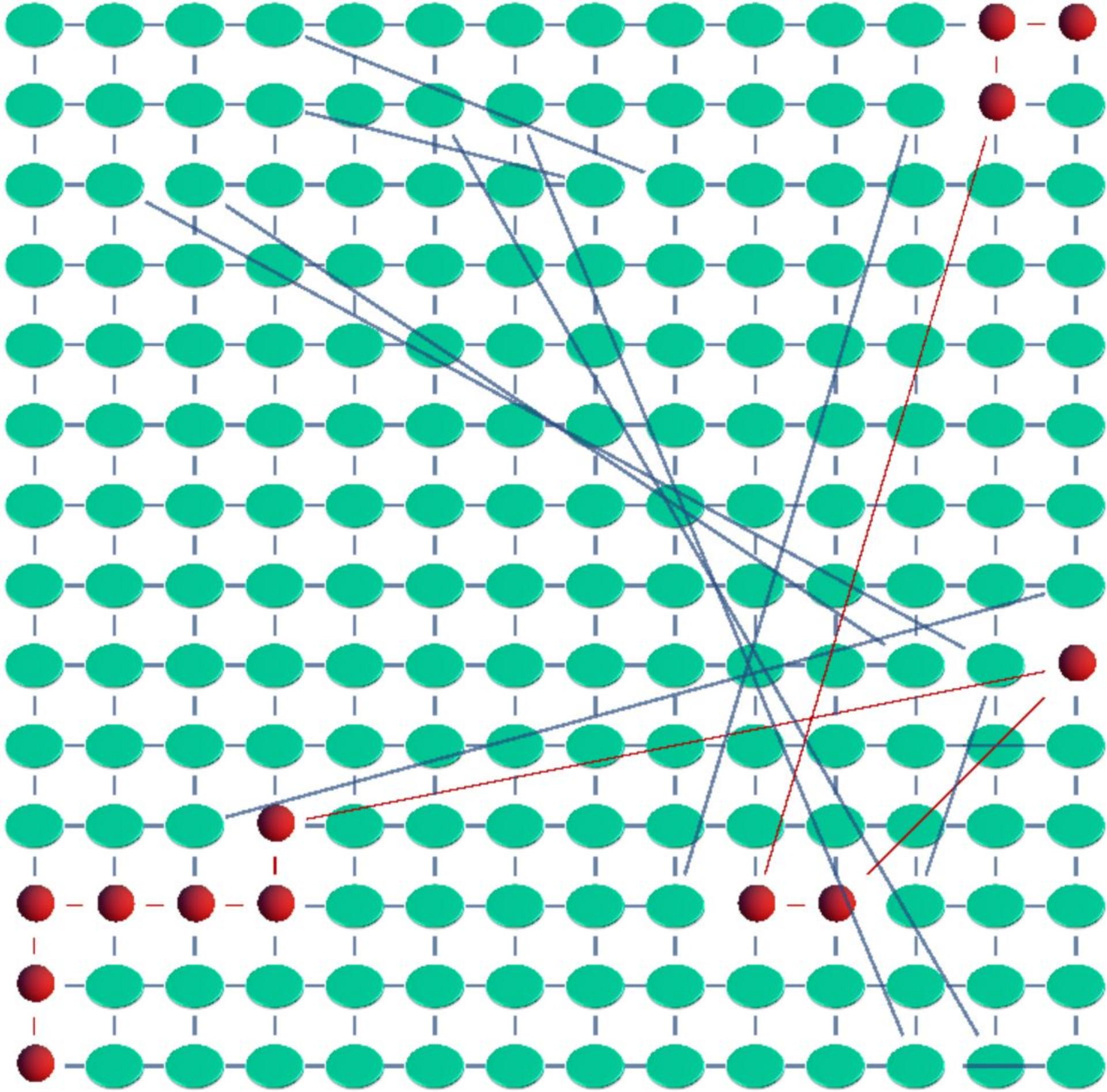}}
\caption{Small world topology of $N=196$ nodes. 
The vertices and links in red show the shortest path connecting 
the opposites sites of the grid where initial excitation and sink are, 
respectively, located.}\label{fig4}
\end{figure}

In this section, we consider another topology, known as small-world (SW) network, that has been extensively investigated
in literature \cite{Latora06} as describing a plethora of realistic complex networks, ranging from world-wide web to protein structures.
In particular, a small-world network is a graph which interpolates between a
regular square lattice and a random graph (Sec. \ref{ran-sca}), i.e. in which the
distance between any two vertices is of the order of that for a
random graph but, at the same time, the concept of neighborhood is
preserved, as for regular lattices. In other words, this graph is
like a square lattice with the introduction of a few long-range edges
creating short-cuts between distant nodes \cite{SW99} -- see Fig. \ref{fig4}.
\begin{figure}[b]
\centerline{\includegraphics[width=.7\textwidth]{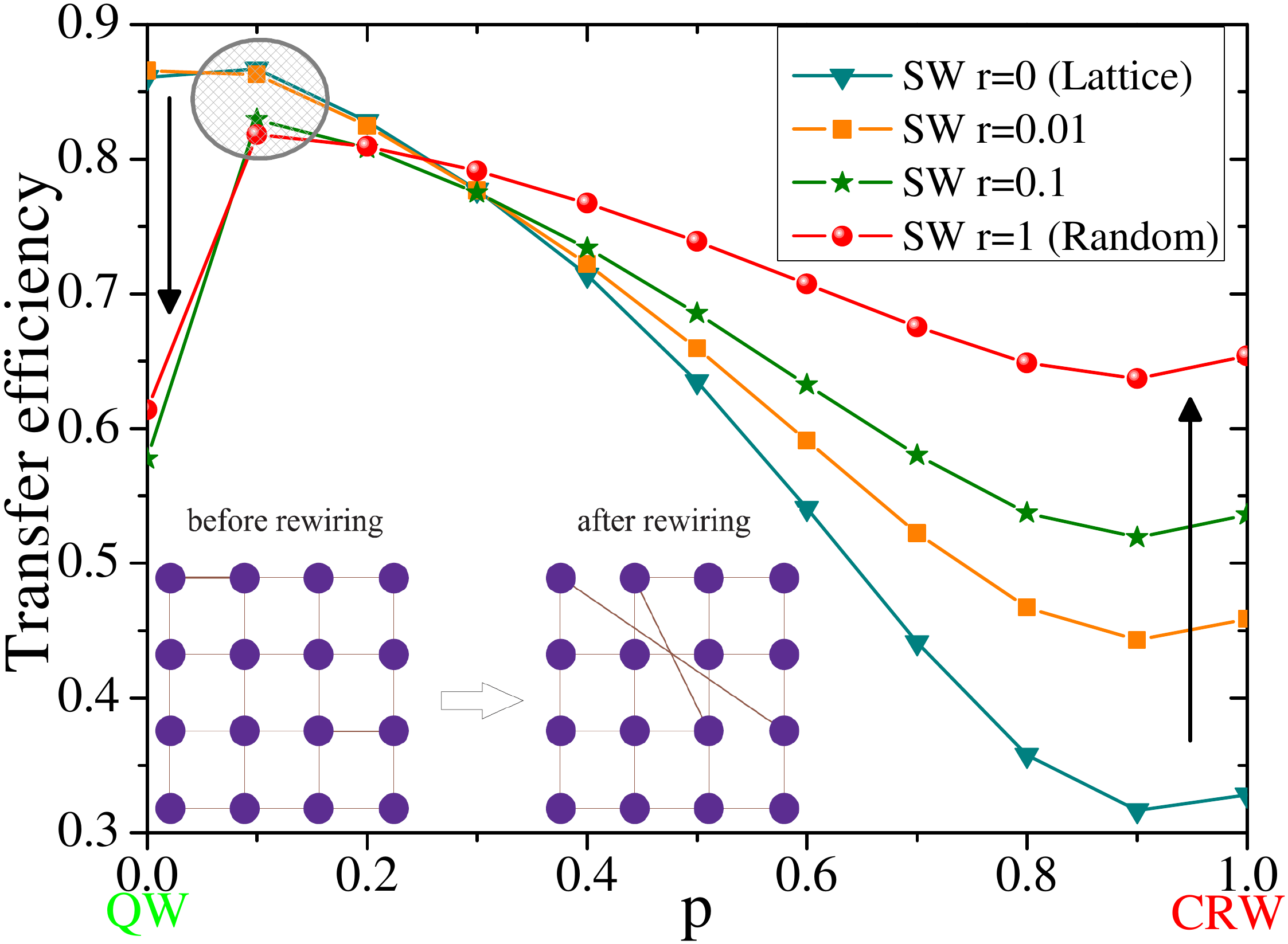}}
\caption{Transfer efficiency ${\cal E}(p)$ as a function of $p$ for the case of small-world networks at different values of the rewiring probability $r$. The arrows are included to point out that the transfer efficiency increases (decreases) when adding long-range links for the classical (quantum) case. Inset: A schematic representation of the rewiring procedure interpolating between a regular ($r=0$) and a random topology ($r=1$) by keeping
fixed and equal to $4$ the degree of each vertex \cite{CLPRT2006}.}\label{fig5}
\end{figure}
\begin{figure}[t]
\centerline{\includegraphics[width=.75\textwidth]{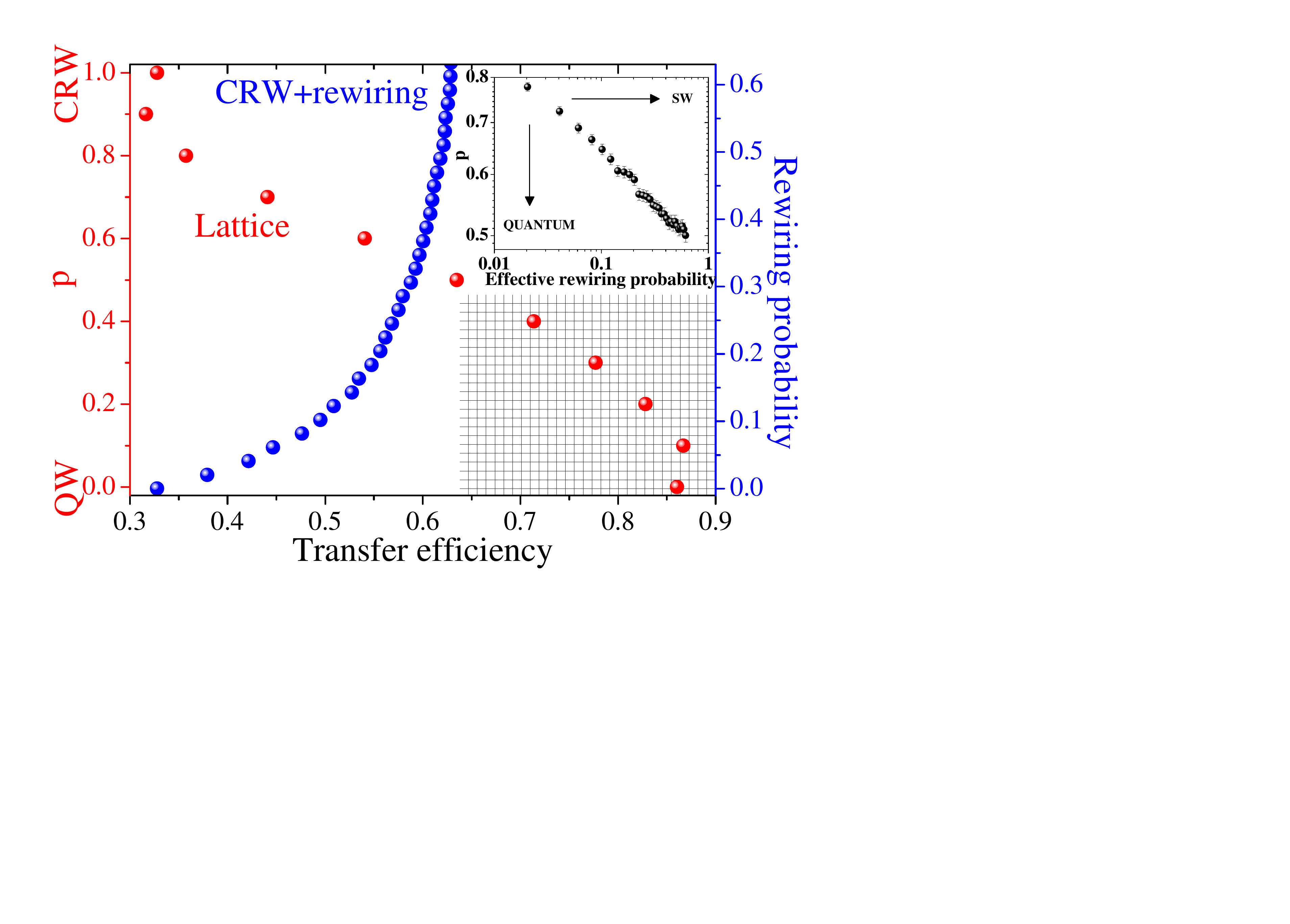}}
\caption{SW transition: QSW transfer efficiency ${\cal E}(p)$ when varying $p$ for a square lattice [$r=0$] (left vertical axis, red) and when changing the rewiring probability $r$ of a small-world network for CRW [$p=1$] (right vertical axis, blue). The grid-shadow region cannot be achieved by CRWs no matter how many long-range links are added on the graph. Inset: Effective rewiring probability $r_e$ for a CRW to reproduce the same transport efficiency of a lattice-QSW with different values of $p$. The presence of more quantum coherence, in the square lattice-QSW case, corresponds to an higher number of long range links in the SW-CRW model and these two quantities scale as a power law (with exponent $\simeq - 0.13$).}\label{fig6}
\end{figure}

It can be constructed following the method proposed in Ref. \cite{SW99}, but here we follow
a slightly different algorithm allowing us to keep fixed the
degree of each vertex as in Ref. \cite{CLPRT2006}.
We start from a regular square lattice and we randomly choose a vertex $V_{i_1}$ and the edge
$V_{i_1}-V_{i_2}$ that connects vertex $V_{i_1}$ to its nearest
neighbor $V_{i_2}$, at random. With probability $r$ this edge is
rewired and with probability $1-r$ it is left in place. If the
edge has been rewired, (a) we choose at random a second vertex
$V_{j_1}$ and one of its edges, e.g. the edge $V_{j_1}-V_{j_2}$
connecting $V_{j_1}$ to $V_{j_2}$, and (b) we replace the couple
of edges $V_{i_1}-V_{i_2}$ and $V_{j_1}-V_{j_2}$ with the couple
$V_{i_1}-V_{j_2}$ and $V_{j_1}-V_{i_2}$, as in the inset of Fig. \ref{fig5}.
This process is repeated by moving over the entire lattice
considering each vertex in turn until one lap is completed. In
such a way the limit case $r=1$ corresponds to a random graph with fixed
degree equal to $4$. In the intermediate cases $0<r<1$ an
increasing number of long-range edges turns up in the graph. In
other terms, the introduction of a few long-range edges create
short-cuts that connect vertices that otherwise would be much
further apart. Strictly speaking, the characteristic
path length $\cal L$ of the rewired graph decreases while increasing $r$ \cite{Latora06}. Therefore,
we can use the term ``small-world'' to refer to a rewired lattice
(with fixed degree) with the minimum number of rewired edges such
that the characteristic path length $\cal L$ is as small as that
one for the corresponding random graph (with $\cal L$ depending
at most logarithmically on the network size $N$), but the cluster coefficient $\cal C$ is still as high as for a regular lattice, i.e. much larger than $\cal C$ for random networks \cite{SW99}. As shown
also in Ref. \cite{CLPRT2006}, this is obtained already
for very small values of $r$ ($r \simeq 0.01$), much before the
random graph limit ($r=1$). Notice that this model undergoes a
`genuine' continuous phase transition as the density of shortcuts
tends to zero with a characteristic length diverging as $r^{-1}$
\cite{Latora06}.
Now we study the QSW dynamics over this geometry when varying the mixing $p$ of classical and quantum behavior. Again, as shown in Fig. \ref{fig5}, we find the optimality of $p \sim 0.1$ for different values of rewiring probabilities $r$. Notice that the curves in Fig. \ref{fig5} become flatter and flatter for higher $r$ because, in the limit of random graph, ${\cal E}(p)$ is expected to weakly depend on $p$.
See also Ref. \cite{MPB2007} for other studies of quantum transport on small-world networks.

\begin{figure}[t]
\centerline{\includegraphics[width=.66\textwidth]{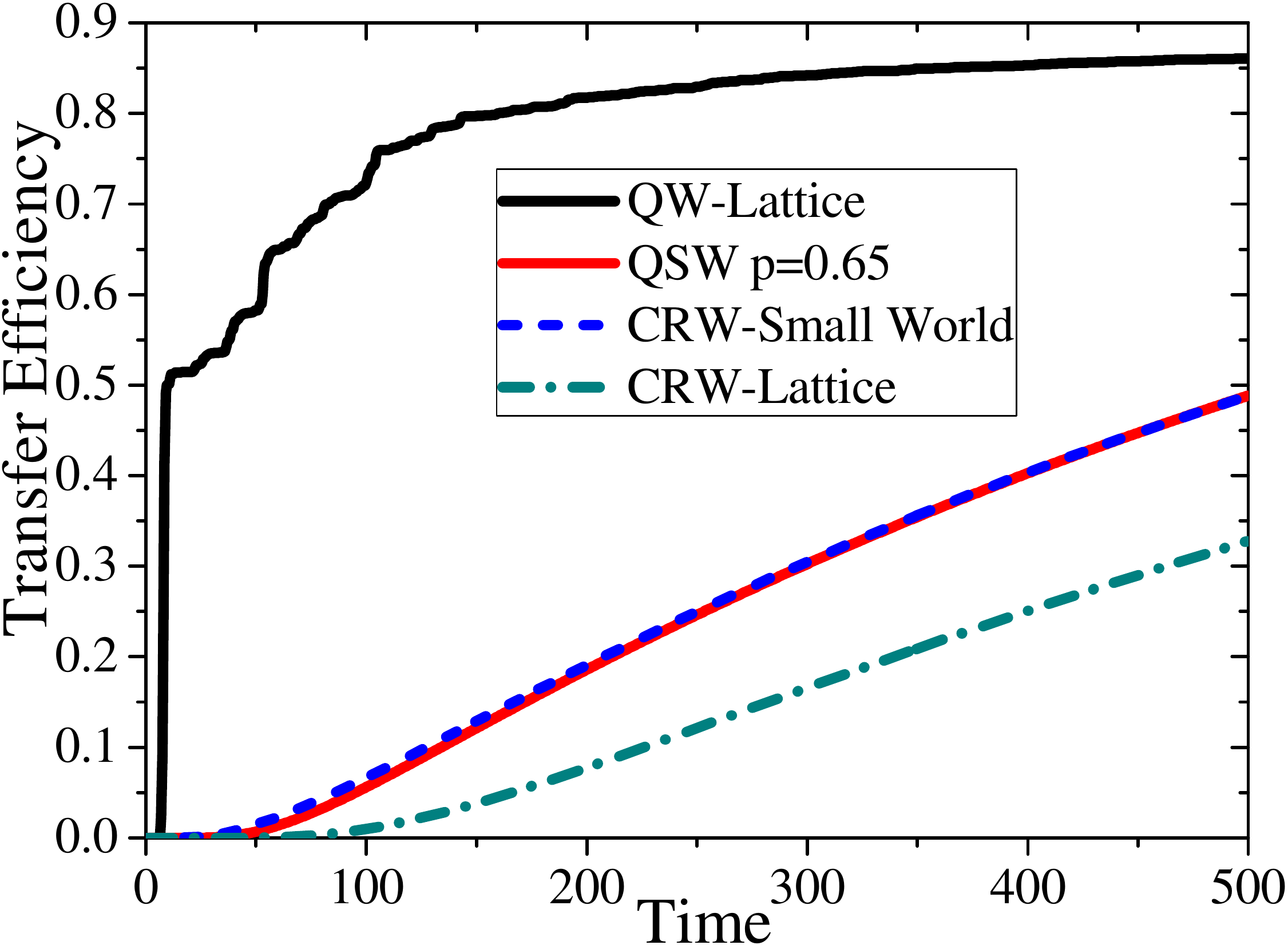}}
\caption{Comparison of ${\cal E}(p,t)$ time evolutions of a CRW ($p=1$) on a SW network with $r \sim 0.09$ and a QSW on a square lattice for $p \sim 0.65$. The extreme cases of QWs ($p=0$) and CRWs ($p=1$) on a square lattice are also shown.}\label{fig7}
\end{figure}
\begin{figure}[b]
\centerline{\includegraphics[width=.6\textwidth]{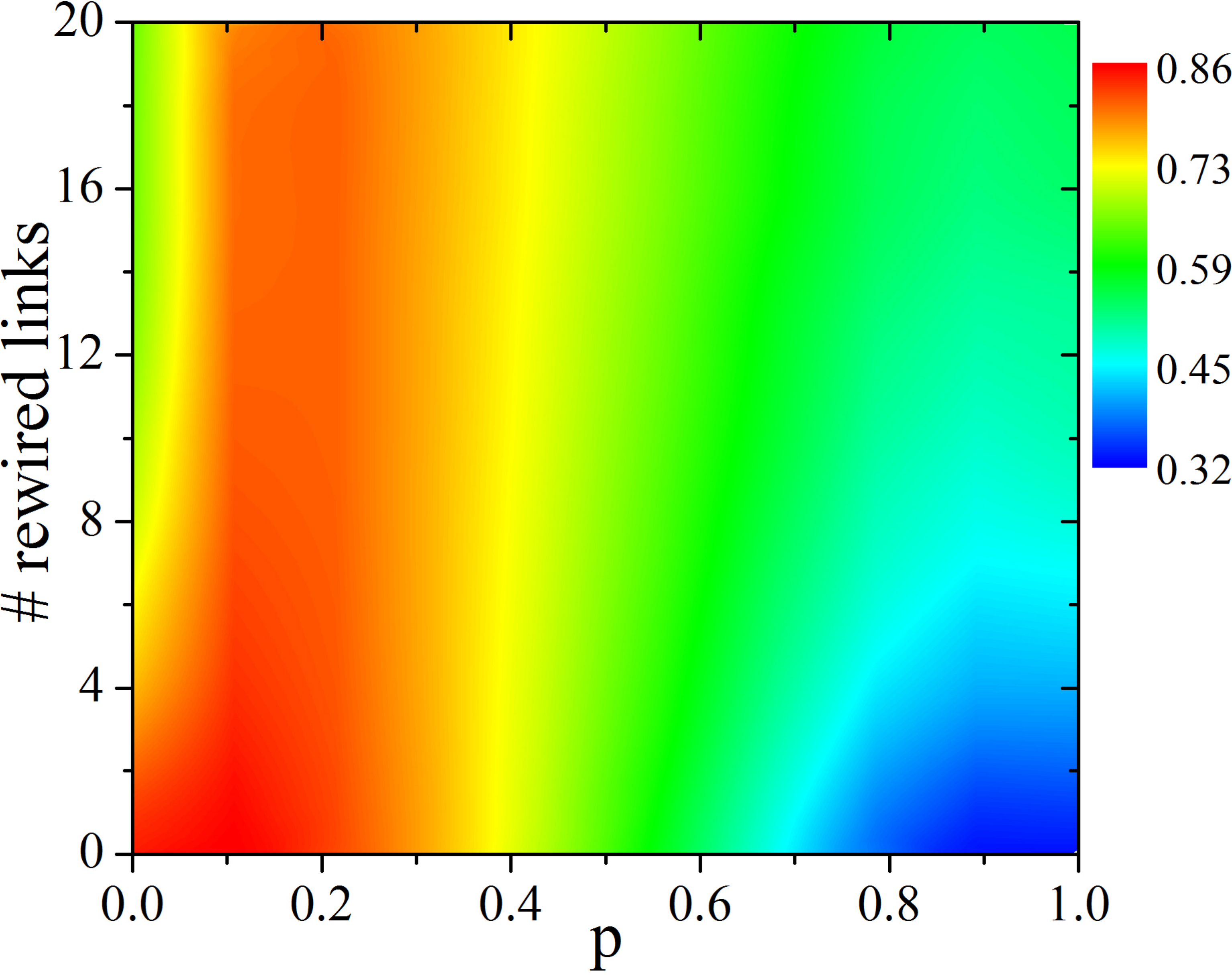}}
\caption{Behaviour of the transfer efficiency ${\cal E}(p)$ vs. the number of randomly rewired links for QSWs on a square lattice geometry of $N=196$ nodes.}\label{fig8}
\end{figure}
Furthermore, we analyze the connection, if any, between the value of $p$ for QSWs on a 
square lattice and the value of $r$ for CRW on a rewired small-world structure (SW-CRW) -- see Fig. \ref{fig6}. 
In other words, we extract the effective value of rewiring probability $r_e$ providing the same transport efficiency corresponding to a given value of $p$. It turns out that these two quantities $p$ and $r_e$ are related and, in particular, follow a power law behavior (inset of Fig. \ref{fig6}). The more coherent is the lattice-QSW dynamics, the higher is the number of long-range links required in the SW-CRW model in order to achieve the same transferred energy into the sink. A comparison of the corresponding time evolutions ${\cal E}(p,t)$ of the transfer efficiency for a given value of $r_e$ and $p$ are shown in Fig. \ref{fig7}. Therefore, not only ${\cal E}(p)$ is the same in correspondence of the mapped $p$ and $r_e$, but also the full time evolutions ${\cal E}(p,t)$ follow the same behaviors. Besides, let us point out that there is a range of high values of transfer efficiency obtained for small $p$ (i.e., closer to the quantum limit), that no CRW can achieve no matter how many long-range links may be added. Examples of kinetic models of energy transport with nonlocal links were studied in Ref. \cite{CS09}.

\begin{figure}[t]
\centerline{\includegraphics[width=.6\textwidth]{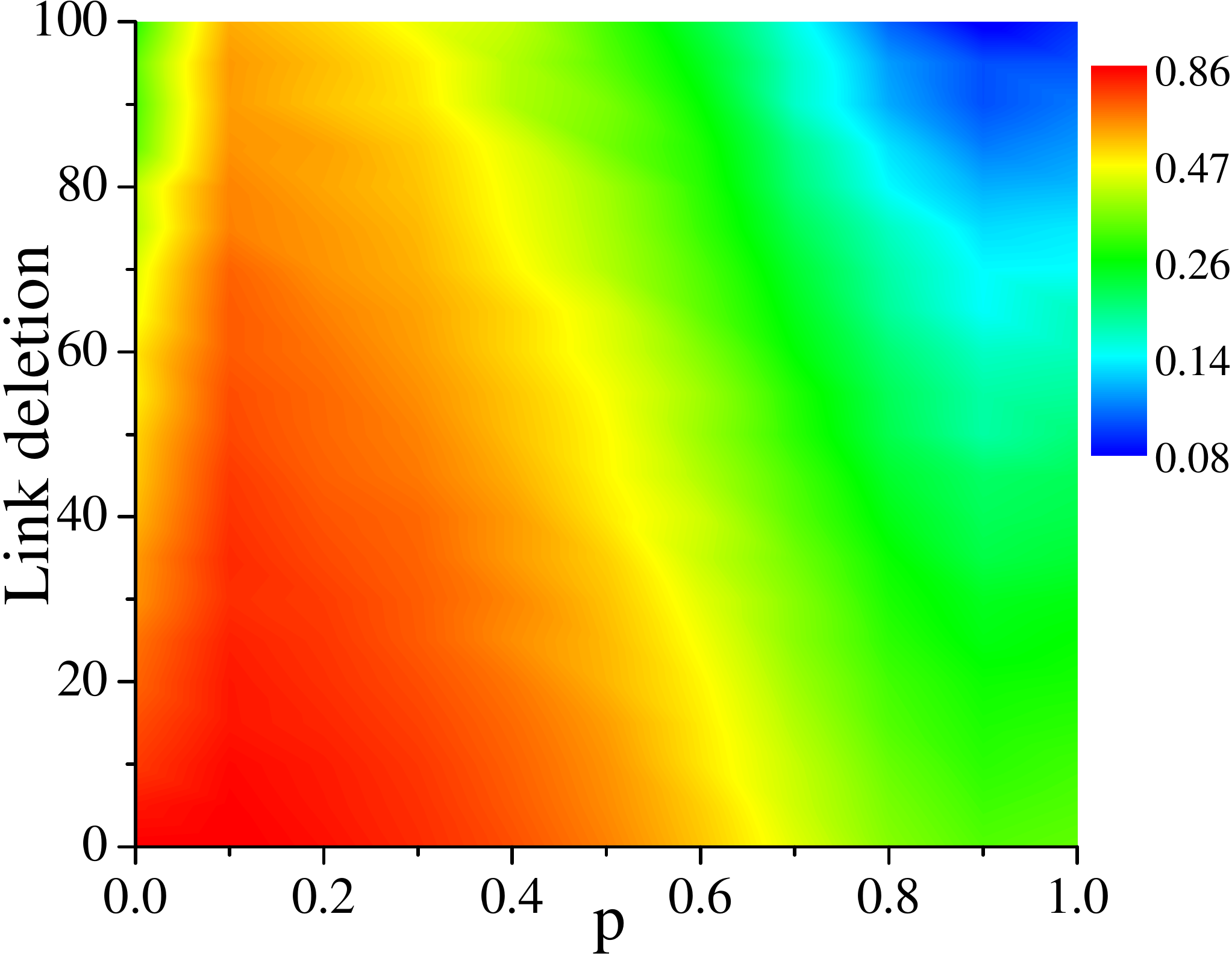}}
\caption{Behaviour of the transfer efficiency ${\cal E}(p)$ vs. the number of randomly deleted links for QSWs on a square lattice geometry of $N=196$ nodes.}\label{fig9}
\end{figure}

Finally, we focus on the robustness of the transport efficiency optimality with respect to rewiring or deleting links. In the context of complex network theory, this is called static robustness and usually refers to the resilience of real graphs (as electric networks, WWW, social networks, etc.) to external attacks or random failures \cite{Latora06}. As shown in Figs. \ref{fig8} and \ref{fig9}, the optimality of $p \sim 0.1$ is very robust against an increasing number of rewired or deleted links. This geometric robustness is also especially relevant when one is dealing with real physical systems behaving as noisy quantum walkers on imperfect structures.
\subsection{Random and Scale-Free graphs}
\label{ran-sca}

\begin{figure}[t]
\minipage{0.55\textwidth}
\centerline{\includegraphics[width=.65\textwidth]{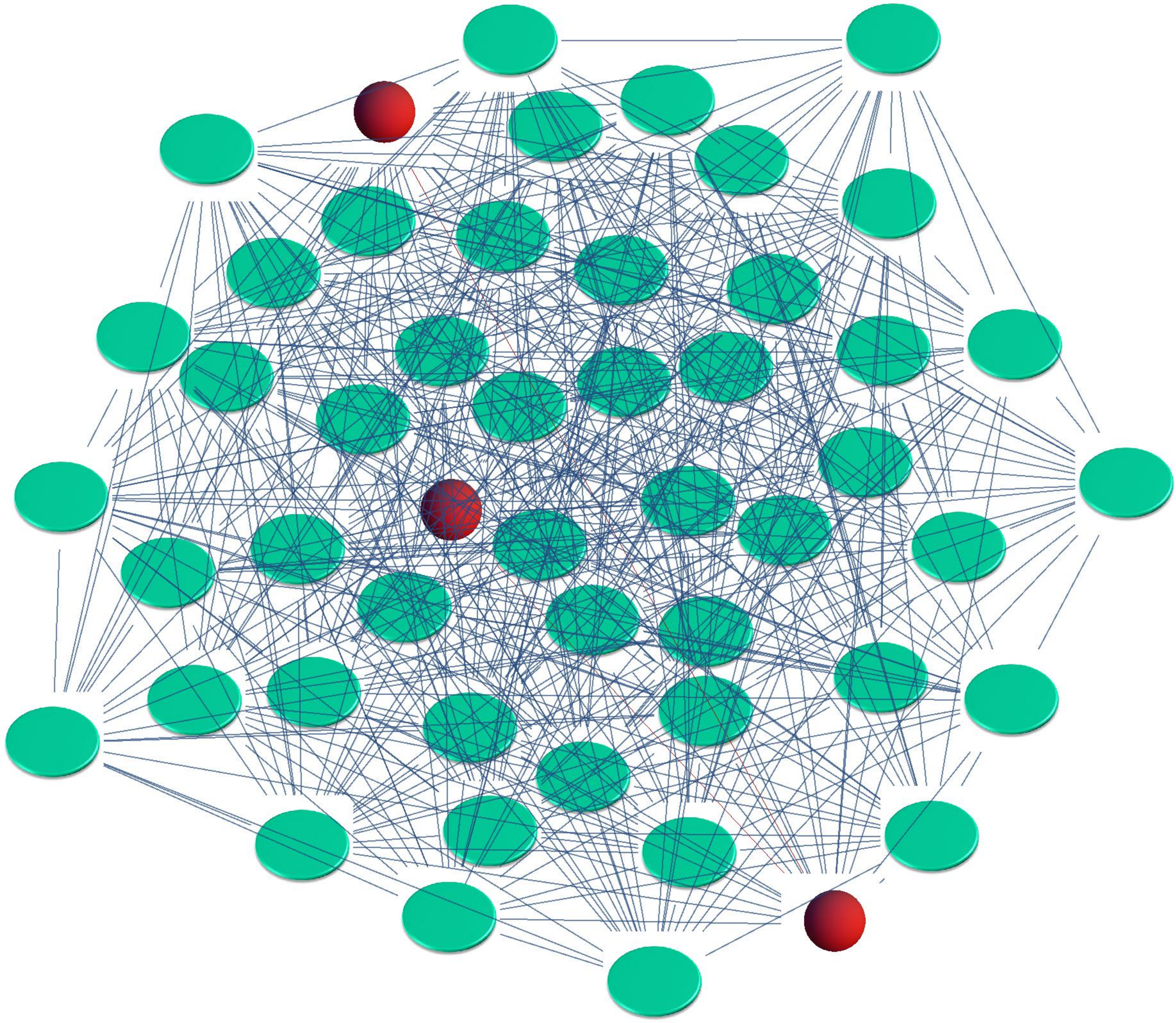}}
\endminipage
\minipage{0.55\textwidth}
\centerline{\includegraphics[width=.65\textwidth]{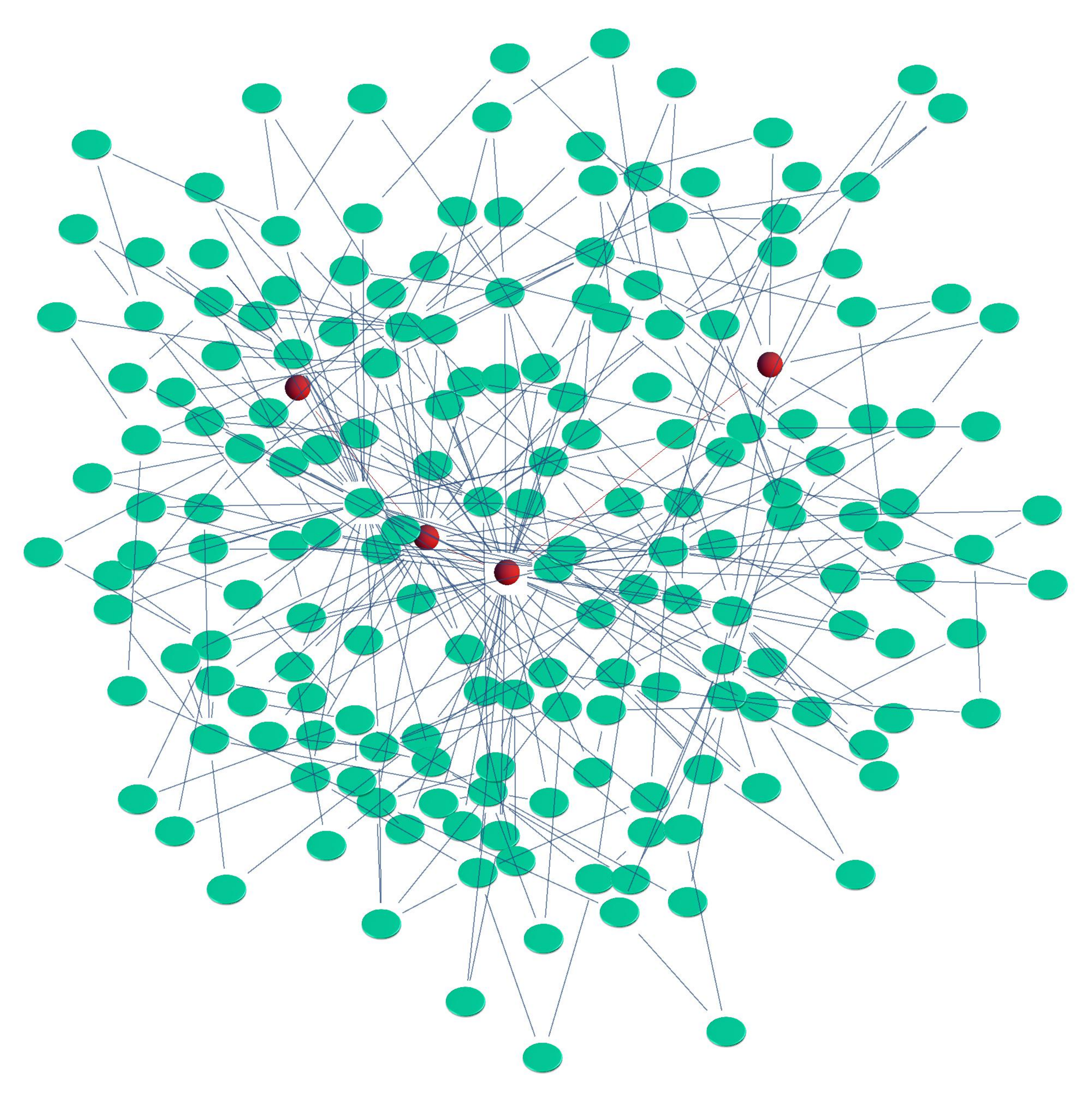}}
\endminipage
\caption{Examples of a RG (left) and a SF network with $\gamma \sim 3$ (right). 
The red nodes represent a geodesic path from the vertex $1$ (initial excitation) to the vertex $N$ which the sink is connected to.}\label{fig10-11}
\end{figure}

Another important class of complex networks is represented by random graphs (RG) and scale-free (SF) networks \cite{Latora06}. 
RGs are defined as structures with Poissonian distribution of the node degree, and can be constructed also as limiting case ($r \rightarrow 1$) of the small-world networks studied above -- see example in the left panel of Fig. \ref{fig10-11}. They are also characterized by small values of both $\cal L$ and $\cal C$. A scale-free network is, instead, a graph with a power-law degree distribution $p(d)\sim d^{-\gamma}$ of the vertex degree $d$. It displays a
small characteristic path length $\cal L$ as for a small-world network and for random graph, but it differs from them for having a power
law degree distribution. This graph can be constructed in the following way. By using the preferential attachment growing procedure
introduced by Barab\'{a}si and Albert \cite{albert}, one starts from $v+1$ all to all connected vertices and at each time step one
adds a new vertex with $v$ edges. These $r$ edges point to old vertices with probability $q_i=\frac{d_i}{\sum_j d_j}$,
where $d_i$ is the degree of the vertex $V_i$, as defined in Sec. \ref{defs}. This procedure allows a selection of the $\gamma$ exponent of the power law degree scaling, with $\gamma=3$ in the thermodynamic limit (i.e., $N \longrightarrow \infty$) -- see example in the right panel of Fig. \ref{fig10-11}.
\begin{figure}[b]
\centerline{\includegraphics[width=.7\textwidth]{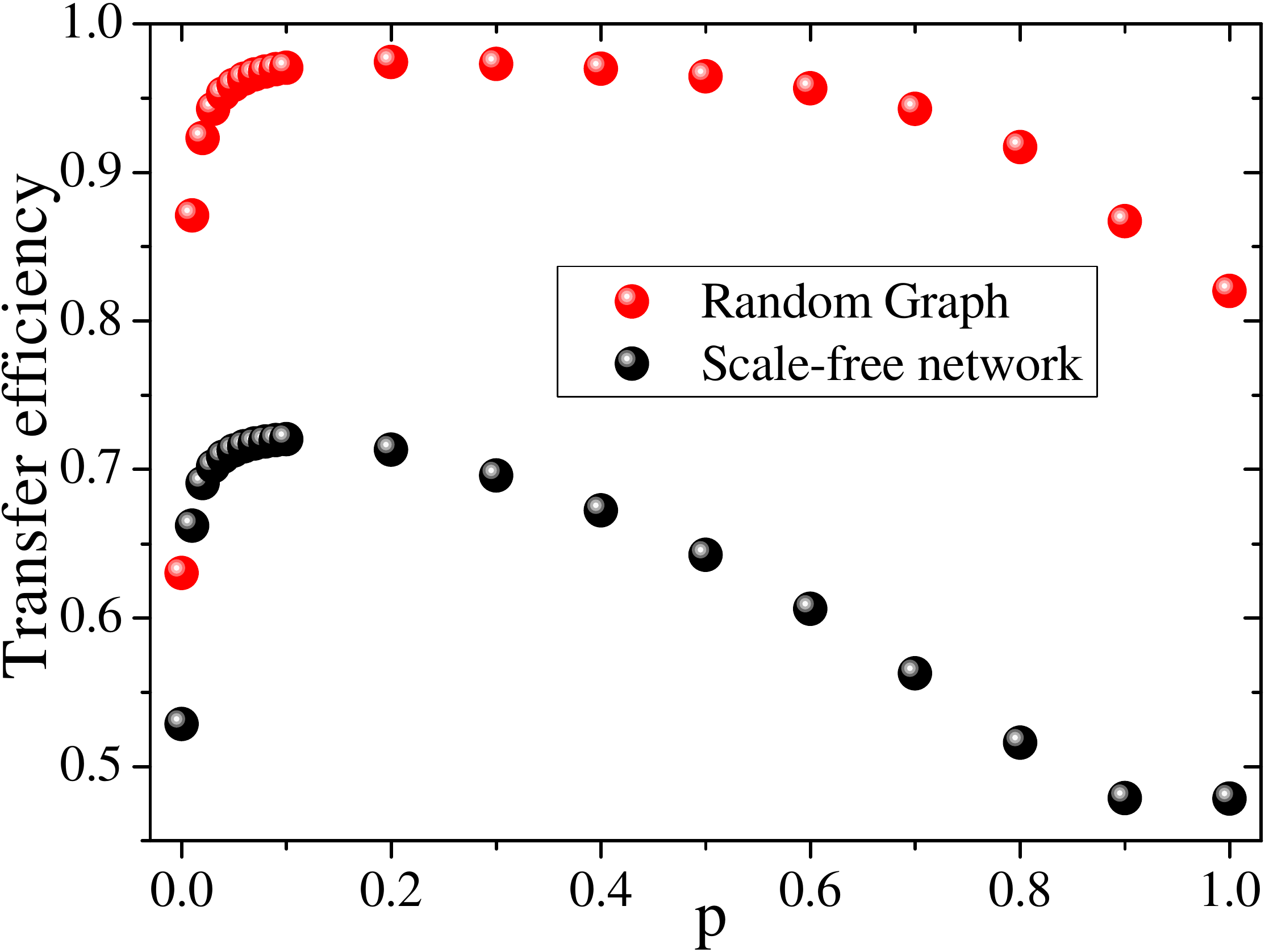}}
\caption{Transfer efficiency ${\cal E}(p)$ as a function of $p$ for the case of RG ($N=200$) and SF network ($N=187$).}\label{fig12}
\end{figure}
The $p$-dependence of the transfer efficiency ${\cal E}(p)$ for both graphs is shown in Fig. \ref{fig12}. Again, but still interestingly enough, the transport efficiency optimality is reached for $p \sim 0.1$.
Another study of continuous-time quantum walks on random graphs, but only for CRWs and QWs, is shown in Ref. \cite{Agliari11}.

\subsection{Other Complex Networks}
\begin{figure}[t]
\centerline{\includegraphics[width=.7\textwidth]{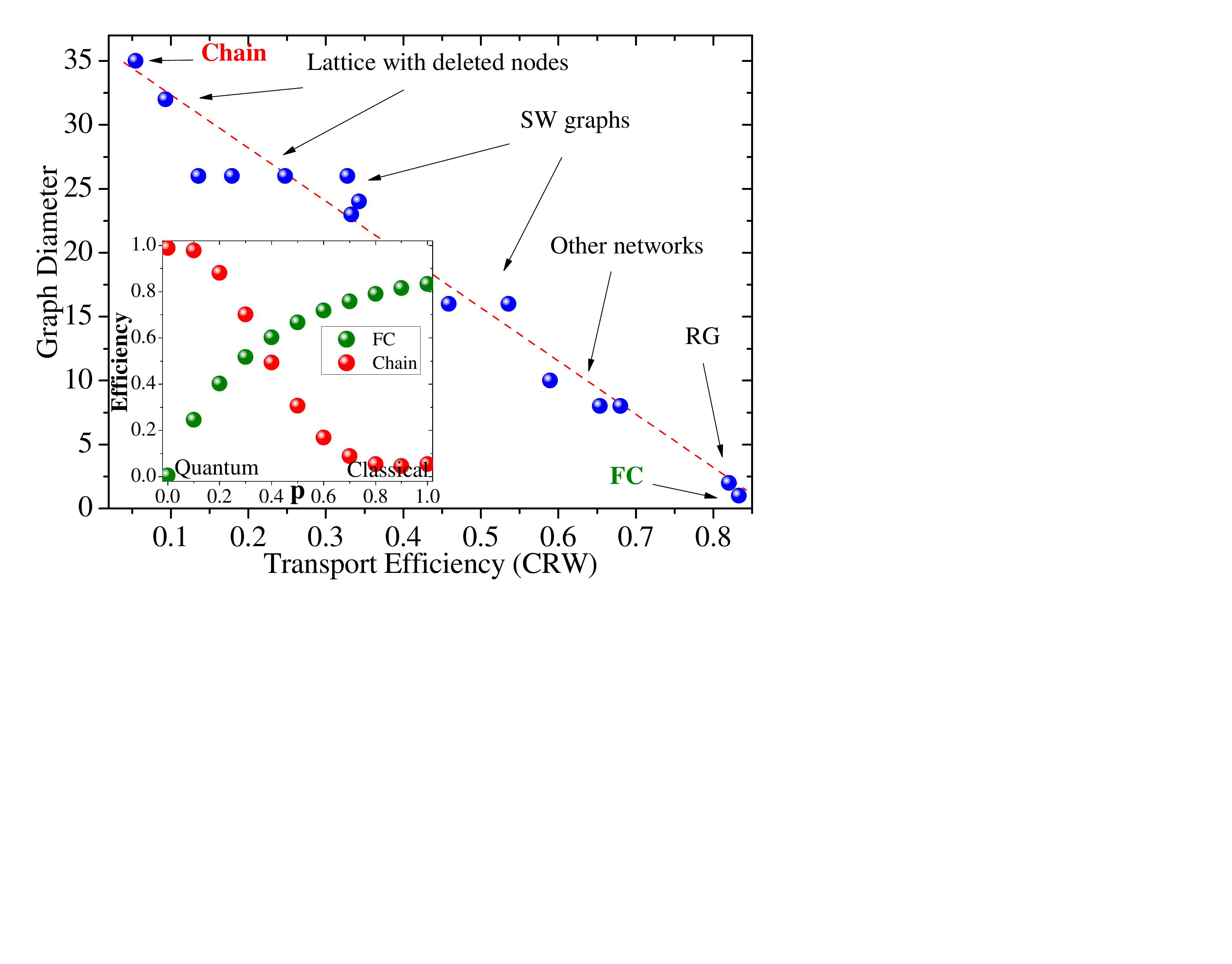}}
\caption{Graph Diameter $D$ as a function of the CRW transport efficiency 
$ {\cal E}(p=1,\bar{t})$ over different complex networks, with $\bar t$ being linearly proportional to the 
number of nodes $N$. 
The sites for the initial excitation and for the sink are randomly chosen.
Inset: ${\cal E}(p)$ vs. of $p$ for the extreme cases of linear chains and FC graphs.}\label{fig13}
\end{figure}
Here we generalize the analysis above for a wider class of large complex networks. First of all, we find that, as expected, the transport efficiency ${\cal E}(p)$ of CRW dynamics is almost linearly dependent on the graph diameter $D$ introduced in the Sec. \ref{defs} -- see Fig. \ref{fig13}. In other terms, the larger $D$ is, the lower is the transport efficiency ${\cal E}(p)$ on the corresponding network. However, it turns out that the latter can be enhanced by adding some quantum coherence ($p<1$) in the dynamics, as found above. Indeed, for a very large class of complex networks (including random graphs, small-worlds, scale-free networks, rings, stars, dendrimers, Kary trees, etc. \cite{Latora06}), with $D > 3$, we find that the optimal transport efficiency is almost always achieved when
\begin{equation}
p_{opt} \sim 0.1 \; .
\end{equation}
The main exception is represented by FC graphs ($D=1$) that are 
optimal for $p=1$, i.e. for CRWs -- see inset of Fig. \ref{fig13}. 
This can be intuitively explained by the fact that for graphs with 
very small $D$ (so high mixing rate $\tau_{mix}$), as FC networks, 
CRWs propagate extremely quickly, while they are very slow for a linear 
chain (largest $D$, so smallest $\tau_{mix}$). Vice versa, a basically 
opposite behaviour is observed for QWs, i.e. larger $D$ usually implies 
faster transport. The physical intuition behind it is that for large $D$ one 
has an high number $\pi$ of different eigenstates of the adjacency matrix 
$A$ (since $\pi \geq D+1$), and so less energy trapped (or localized) states 
in the dynamics. This can be better understood by means of the notion of 
invariant subspaces introduced in Ref. \cite{CCDHP2009}. The latter are defined as set of eigenstates of
the Hamiltonian that are orthogonal with the site connected to the sink, i.e. 
they are not affected by the open-system dynamics and their evolution is purely coherent and described by just a global phase. 
These invariant (trapped) states can be systematically found for any network 
with some degeneracy, as geometrical symmetries corresponding to large degenerate eigenspaces, and the more they are, the 
larger is the amount of energy trapped in the system. Hence, a FC graph ($D=1$ and $\pi=2$) is the worst geometry for QWs; 
indeed, because of destructive interference effects and the presence of as large as possible invariant subspace, the transfer efficiency ${\cal E}(p=0)$  
cannot be larger than $1/(N-1)$ \cite{CCDHP2009}. However, by means of a universal mixing 
($p \sim 0.1$) of, loosely speaking, $90\%$ of QWs and $10\%$ of CRWs, the invariant (trapped) subspaces 
are destroyed and the energy transport becomes optimal and robust, and, in particular, we find that 
this is irrespective of the particular underlying geometry of the complex network. 

\section{Conclusions and Outlook}
\label{conclu}

In this paper we extensively investigate the transport properties of quantum stochastic walks over a wide family of large complex networks. More specifically, we numerically calculate and compare the transport performances of these graphs by the transfer efficiency of a trapping site, also motivated by the structure of light-harvesting complexes where the interplay of quantum coherence and environmental noise has been recently shown to play a fundamental role in explaining the remarkably efficient as well fast exciton energy transfer. We find that, roughly speaking, the mixing of $90\%$ of quantum dynamics and $10\%$ of classical one, i.e. for $p \sim 0.1$, leads to optimal transport efficiency. 
On top of that, this optimal behaviour is not only relatively common on very different networks 
but also robust with respect to geometric changes as rewiring or deleting links.

However, there are some exceptions represented by very well connected graphs 
(e.g., FCs and graphs with $D \ll N$), where instead classical random walks, 
i.e. $p=1$, provide the optimal transport into the trapping site. 
This might be a consequence of the fact that for such graphs there are many trapped states due to destructive interference (large invariant subspaces almost covering the full Hilbert space \cite{CCDHP2009}), that are fully destroyed only at the classical limit. The transition behaviour from $p=1$ to $p \sim 0.1$ when making the network less and less connected ($D \gg 1$), and a deeper analysis of this widespread and robust optimality, also based on a possible analytical derivation by means of Lieb-Robinson bounds \cite{eisert13a,eisert13b}, will be investigated in a forthcoming paper.

Finally, our results could be tested through already experimentally available benchmark platforms as cold atoms in optical lattices, artificial light-harvesting structures, and photonics-based architectures, hence also inspiring the design of optimally efficient transport nanostructures for novel solar energy and information quantum-based technologies.

\subsection*{Acknowledgments}

This work has been supported by EU FP7 Marie--Curie Programme (Career Integration Grant) and by a MIUR--FIRB grant (Project No. RBFR10M3SB). We acknowledge QSTAR for computational resources, based also on GPU-CUDA programming by NVIDIA Tesla C2075 GPU computing processors.
\newline
\


\begin{thebibliography}{99}
\bibitem{note}
A more realistic model including static disorder in both energy and coupling terms, with also specific distance-dependent couplings referred to particular physical systems, will be analyzed in a forthcoming paper.
\end{thebibliography}
\end{document}